\documentclass[12pt]{article}
\usepackage{amsmath,epsf,amssymb,latexsym,cite,shadow,eucal}
\usepackage[ps,dvips,matrix,arrow,frame,import,curve,color]{xy}

\newtheorem{theorem}{Theorem}

\newtheorem{conjecture}{Conjecture}

\newif\iffigs\figstrue

\DeclareFontFamily{U}{rsf}{}
\DeclareFontShape{U}{rsf}{m}{n}{
  <5> <6> rsfs5 <7> <8> <9> rsfs7 <10-> rsfs10}{}
\DeclareMathAlphabet\Scr{U}{rsf}{m}{n}

\def\O{\Scr{O}}
\def\C{{\mathbb C}}
\def\P{{\mathbb P}}

\def\Z{{\mathbb Z}}

\def\NN{{\mathbb N}}

\def\Hom{\operatorname{Hom}}

\def\Ext{\operatorname{Ext}}

\def\Spec{\operatorname{Spec}}
\def\Proj{\operatorname{Proj}}

\def\Tr{\operatorname{Tr}}

\def\GU{\operatorname{U{}}}

\def\id{{\mathbf{1}}}

\def\CY{Calabi--Yau}
\def\LG{Landau--Ginzburg}

\def\cA{{\Scr A}}
\def\cB{{\Scr B}}

\def\DC{\mathbf{D}}
\def\Ch{\mathrm{Ch}}

\def\mf#1{\mathfrak{#1}}

\def\DSing{\operatorname{\mathbf{D}^{\mathrm{gr}}_{\mathrm{Sg}}}}

\begin{document}

\begin{titlepage}
\begin{flushright}
DUKE-CGTP-07-01\\
hep-th/0703279\\
March 2007
\end{flushright}
\vspace{.5cm}
\begin{center}
\baselineskip=16pt
{\fontfamily{ptm}\selectfont\bfseries\huge
Topological D-Branes\\[3mm]
and Commutative Algebra}\\[20mm]
{\bf\large  Paul S.~Aspinwall
 } \\[7mm]

{\small

Center for Geometry and Theoretical Physics, 
  Box 90318 \\ Duke University, 
 Durham, NC 27708-0318 \\ \vspace{6pt}

 }

\end{center}

\begin{center}
{\bf Abstract}
\end{center}
We show that questions concerning the topological B-model on a
Calabi--Yau manifold in the Landau--Ginzburg phase can be rephrased in
the language of commutative algebra. This yields interesting and very
practical methods for analyzing the model. We demonstrate how the
relevant ``Ext'' groups and superpotentials can be computed
efficiently by computer algebra packages such as Macaulay. 
This picture leads us to conjecture a general description of
D-branes in linear sigma models in terms of triangulated
categories. Each phase of the linear sigma model is associated with a
different presentation of the category of D-branes.
\vspace{2mm} \vfill \hrule width
3.cm \vspace{1mm} {\footnotesize \noindent email: psa@cgtp.duke.edu}
\end{titlepage}

\vfil\break


\section{Introduction}    \label{s:intro}

As is well-known, the topological A-model is associated to the symplectic
geometry of a \CY\ manifold, while the B-model is associated to
algebraic geometry. This, in addition to the fact that the A-model
suffers from instanton corrections, makes the A-model ``difficult'' and
the B-model ``easy''. But just how easy is the B-model? Suppose our
\CY\ manifold is a hypersurface in $\P^n$ given by a constraint $W=0$.
In this case, the spectrum of closed strings and the structure of their
correlation functions is given by the chiral ring \cite{VW:}
\begin{equation}
   \frac{k[x_0\ldots x_{n-1}]}{(\partial_0 W,\ldots,\partial_{n-1}W)}.
\end{equation}

For open strings we need to work harder. First we need to classify the
boundary conditions, or D-branes. This is given by the (bounded)
derived category of coherent sheaves on $X$ (see, for example
\cite{Kon:mir,Shrp:DC,Doug:DC,AL:DC}).
Then, given two D-branes $E$ and $F$, we may compute the Hilbert space
of open strings between these D-branes from $\Ext^k(E,F)$. Certain
correlation functions between these open strings form an
$A_\infty$-structure as analyzed in \cite{W:CS,HLW:Ainf,AK:ainf}.
These correlation functions are associated to the superpotential of an
associated D-brane world-volume field theory.

A principal idea in our analysis is the use of matrix factorizations
to represent the D-branes. Such matrix factorizations are known to be
a natural language for D-branes in the Landau--Ginzburg phase of a
\CY\ compactification \cite{KL:Mfac,KL:TLG1,AADia:GepD}. The
equivalence of matrix factorizations and coherent sheaves was
discussed in
\cite{AADia:GepD,Brunner:2004mt,DGJT:D,Hori:2004ja,HW:mfac,
  Brunner:2005fv,Enger:2005jk} and then proven in \cite{Orlov:mfc}.
The explicit form of the map between these categories was analyzed in
\cite{me:lgdict}.  Confronted directly, the algebra of matrix
factorizations seems very unwieldy but in this paper we will show that
all the necessary calculations can be reduced to questions in
commutative algebra.  Furthermore, commutative algebra packages such
as Macaulay \cite{M2:M2} are well-suited to performing the necessary
computations.  Indeed, the method that Macaulay uses to compute Ext
groups, as discussed in \cite{AG:McExt}, is uncannily well suited to
our needs.

As is often the case, a better understanding of computations will give
further insight into the basic structure at hand. We will see that it
is natural to consider matrix factorizations of the superpotential
that appears in the gauged linear $\sigma$-model rather than the
superpotential of the \LG\ theory. The category of D-branes will then
be a quotient of this matrix factorization category. Which quotient is
taken depends on which ``phase'' of the linear $\sigma$-model we are
in. This leads to a conjectured equivalence between many different
triangulated categories.

In section \ref{s:mfac} we will discuss topological D-branes from the
point of view of matrix factorizations and, in particular, we will tackle the
problem of how best to compute the Hilbert spaces of open strings
between such D-branes. We believe the best way to understand this
analysis is through a practical computation using Macaulay and so we
give the details of a specific example.

In section \ref{s:superP} we discuss the $A_\infty$-structure of the
D-brane category. This again can be computed practically using
Macaulay. We also see explicitly how to map from closed string
vertex operators to open string boundary-preserving operators.

In section \ref{s:lin} we see that the construction of section
\ref{s:mfac} naturally lends itself to a general conjecture of how to
construct a category of D-branes in any phase of the gauged linear
$\sigma$-model. We show that this conjecture works in many cases.
Finally in section \ref{s:conc} we present some conclusions.


\section{Matrix Factorizations}  \label{s:mfac}

\subsection{The D-Brane category} \label{ss:Dbrane}

Let us review the construction of Kapustin and Li
\cite{KL:Mfac}. Consider a \LG\ theory in $k^n$ with coordinates
$x_0.,\ldots, x_{n-1}$ and superpotential
$W(x_0,\ldots,x_{n-1})$. Physically we will want the field $k$ to be
equal to the complex numbers, $\C$. However, for computational
efficiency, we will also want to consider other fields.

Let $B$ denote the polynomial ring $k[x_0,\ldots,x_{n-1}]$. We
construct the D-brane category as follows.  Objects $\bar P$ are
ordered pairs of free $B$-modules of arbitrary but equal finite rank with
maps between them going in each direction:
\begin{equation}
  \bar P = \Bigl(
\xymatrix@1{
  P_1 \ar@<0.6mm>[r]^{p_1}&P_0\ar@<0.6mm>[l]^{p_0}
}\Bigr). \label{eq:Pdef}
\end{equation}
The two maps satisfy the matrix factorization condition
\begin{equation}
  p_0p_1 = p_1p_0 = W.\id.   \label{eq:mfact}
\end{equation}

Trivially we may also combine $p_0$ and $p_1$ to form a map
\begin{equation}
  p: P_0\oplus P_1 \to P_0\oplus P_1,
\end{equation}
where
\begin{equation}
  p^2 = W.\id.   \label{eq:mfact2}
\end{equation}

A map $f:\bar P\to\bar Q$ is simply a pair of maps $f_0:P_0\to Q_0$
and $f_1:P_1\to Q_1$ such that all squares commute. 
A map is said to be null-homotopic (or a null-homotopy) if there are
maps $s_0:P_0\to Q_1$ and $s_1:P_1\to Q_0$ such that
\begin{equation}
  f_0 = s_1p_0 + q_1s_0,\quad f_1 = q_0s_1 + s_0p_1.
\end{equation}
The category of D-branes is given by the homotopy category obtained by
identifying morphisms with maps modulo null-homotopies.  The Hilbert
space of open strings in the topological B-model between two branes is
given by the space of morphisms in this category.  Note that if either
$p_0$ or $p_1$ is the identity map then the identity map $\bar
P\to\bar P$ is a null-homotopy and thus $\bar P$ is equivalent to 0 in
this category.

As is well-known, for a correspondence with \CY\ manifolds, one must
consider {\em orbifolds\/} of D-branes in a \LG\ theory. For
simplicity let us assume, for now, that the orbifold group is cyclic
$G\cong\Z_d$. This group acts diagonally on the coordinates
$x_0,\ldots, x_{n-1}$ where the eigenvalues give $B$ the structure of
a graded ring. For example, the quintic threefold corresponds to a
\LG\ theory with $n=d=5$. In this case the 5 coordinates have equal
eigenvalues and so one may declare the grade of each $x_i$ to be one.
For a weighted projective space $\P^N_{\{w_0,w_1,w_2,\ldots,\}}$, one
may use $w_i$ as the grade of $x_i$ for the corresponding graded ring
$B$. 

Using this grading structure, we now declare that the two free modules
on $\bar P$ are graded B-modules. Let us define a morphism between two graded
modules as a module map of degree 0. Since $W$ has degree $d$ with
respect to the grading, we cannot declare both $p_0$ and $p_1$ in
(\ref{eq:Pdef}) to be morphisms. Instead we may choose $p_0$ to be a
map of degree $d$, and $p_1$ to be a map of degree 0. Thus, $p_0$ is a
morphism
\begin{equation}
  p_0: P_0 \to P_1(d),
\end{equation}
where $(d)$ denotes a shift in the grading of a module, i.e., $M(d)_m
= M_{d+m}$.

We may define the category GrPair($W$) whose objects are pairs $\bar
P$ of this form and whose morphisms are maps $f:\bar P\to\bar Q$ such
that all squares commute.

If we then identify morphisms which differ by null-homotopies then we
obtain the required D-brane category. Following \cite{Orlov:mfc} we
denote this category $\mathrm{DGr}B(W)$.

As well as the grading, we also have the notion of a ghost number
charge for each object. A shift in this charge will be denoted by square
brackets $[\ldots]$. As is well-established (see \cite{me:TASI-D} for
a review) this shift gives the category of D-branes the structure of a
{\em triangulated category\/}. It was shown in \cite{Wal:LGstab} that
this shift functor acts on $\mathrm{DGr}B(W)$ as
\begin{equation}
  \bar P[1] = \Bigl(
\xymatrix@1{
  P_0 \ar@<0.6mm>[r]^{p_0}&P_1(d)\ar@<0.6mm>[l]^{p_1}
}\Bigr). \label{eq:shift1}
\end{equation}
It follows that
\begin{equation}
  \bar P[2] = \bar P(d). \label{eq:2=d}
\end{equation}

In this paper we will be dealing with many categories. Therefore, any
mention of $\Hom$ or $\Ext$ should properly come with a subscript to
denote the relevant category. We will use the convention that
$\Ext^j(\mathsf{a}, \mathsf{b})=\Hom(\mathsf{a}, \mathsf{b}[j])$
without any subscript refers to the category of D-branes. That is,
$\Ext^j(\mathsf{a}, \mathsf{b})$ is the Hilbert space of open strings
from D-brane $\mathsf{a}$ to D-brane $\mathsf{b}$ shifted by ghost
number $j$.


\subsection{Construction of Resolutions} \label{s:resn}

The above construction of the category $\mathrm{DGr}B(W)$ is
straight-forward but, as defined, it is far from convenient to perform
any computations. In this section we will review a construction
of Avramov and Grayson \cite{AG:McExt}\footnote{In order to retain the
same notation as \cite{Orlov:mfc} we have interchanged the r\^oles of
$A$ and $B$ in this reference.} based on \cite{Eis:mf,Sham:mf}
that turns out to be very well suited to our problem.

Again we let $B$ denote a polynomial ring $k[x_0,\ldots,x_{n-1}]$.
For some of this paper it will suffice to assume that $B$ is singly
graded. For generality, however, we will allow $B$ to be a
multiply graded ring. Let $I$ be a homogeneous ideal
$(f_1,f_2,\ldots,f_c)$ in $B$. We assume that $f_1,f_2,\ldots,f_c$
form a {\em regular sequence\/}. Now define the graded ring $A=B/I$. 

If $B$ is singly and equally graded, then the condition of a regular
sequence is equivalent to the statement that the projective variety
defined by $f_1=f_2=\ldots=f_c=0$ in $\P^{n-1}$ is a complete
intersection. Let $X=\Proj A$ denote this complete intersection.  More
generally we would define $X$ as a complete intersection in a toric
variety.

Consider a finitely-generated graded $A$-module $M$. Equivalently $M$
may be regarded as a $B$-module which is annihilated by $I$. We wish
to study the question of how to compute a free resolution of the
$A$-module $M$:
\begin{equation}
\xymatrix@1{
\cdots\ar[r]&F_2\ar[r]&F_1\ar[r]&F_0\ar[r]&M\ar[r]&0,}
\end{equation}
where $F_j$ is a free module $A(a_{j,0})\oplus A(a_{j,1})\oplus\ldots$
and $a_{j,s}$ are integers denoting ``twists'' of $A$.\footnote{This
  notation is for a singly graded ring. The multiple grading case will
  require multiple twists for each summand.} This will allow
computations of $\Ext_A$ groups.

$\Spec A$ is the affine variety defined by $f_1=f_2=\ldots=f_c=0$ in
$\C^n$. If $\Spec A$ is smooth then we are guaranteed to have a finite
free resolution. It turns out that the modules of interest to us are
the ones with no finite free resolution. We are interested in the case
where $\Spec A$ has an isolated singularity at the origin, and hence $X$
is smooth.

Since $\C^n$ is trivially smooth, as a $B$-module $M$ will have a
finite free resolution
\begin{equation}
\xymatrix@1{
\cdots\ar[r]^{d_C}&C_2\ar[r]^{d_C}&C_1\ar[r]^{d_C}&C_0\ar[r]^r&M\ar[r]&0,} 
   \label{eq:Cres}
\end{equation}
where each $C_i$ is a finitely-generated free $B$-module.

Define the finitely-generated free $B$-module
\begin{equation}
  C = \bigoplus_i C_i.   \label{eq:C}
\end{equation}
Suppose $A$ is multiply graded by $r$ integers. Then $C$ has $r+1$
gradings. The extra grading is given by the subscripts in
(\ref{eq:C}). We will refer to this latter grading as the {\em
  homological grading}. The other $r$ gradings will be called the {\em
  ring gradings}. The map $d_C$ in (\ref{eq:Cres}) then has
homological degree $-1$ and all the ring degrees are 0.

Recall that $c$ denotes the number of equations defining our complete
intersection. Introduce $c$ variables $X_1,\ldots, X_c$ each with a
homological degree of $-2$ and ring degrees of $-\deg f_i$. Now define
the multiply graded ring
\begin{equation}
  S = B[X_1,X_2,\ldots,X_c].
\end{equation}

Set $D=\Hom_B(S,B)$, the dual of $S$. $D$ is naturally endowed
with the structure of an algebra known as the {\em divided powers
  algebra}. We refer to appendix A2 of \cite{Eis:CA} for more
details. We will not need any particular knowledge of this divided
powers structure here.

Let $\gamma$ be a multi-index $(\gamma_1,\ldots,\gamma_c)$
in\footnote{$\NN$ contains 0.} $\NN^c$
and denote the monomial $X_1^{\gamma_1}X_2^{\gamma_2}\ldots$ by
$X^\gamma$. Let $|\gamma|=\sum_i\gamma_i$. Denote the origin
$(0,0,\ldots)$ by $o$ and a basis vector $(0,0,\ldots,1,\ldots,0)$ by
$\epsilon_i$. We copy the following theorem from
\cite{AG:McExt}:

\begin{theorem} \label{th:mf}
  There exists a set of homogeneous $B$-linear maps
\begin{equation}
  d_\gamma:C\to C,
\end{equation}
for all $\gamma\in \NN^c$ with homological degree $2|\gamma|-1$ and
ring degree 0, such that
\begin{equation}
\begin{split}
  d_o &= d_C\\
  \{d_o,d_\gamma\} &= \begin{cases}
      -f_i.\id_C & \text{if $\gamma=\epsilon_i$, $i=1,\ldots,c$}\\[3mm]
      \displaystyle{-\sum_{\substack{\alpha+\beta=\gamma\\ \alpha,\beta\neq o}}
         d_\alpha d_\beta}&\text{otherwise.}
\end{cases}
\end{split}
\end{equation}

These induce a $B$-linear map of homological degree $-1$ and ring
degree 0
\begin{equation}
  d_{CD}: C\otimes_B D \to C\otimes_B D,
\end{equation}
given by
\begin{equation}
  d_{CD}(x\otimes y) = \sum_{\gamma\in \NN^c} d_\gamma(x)\otimes 
     (X^\gamma\lrcorner\,y). \label{eq:dCD}
\end{equation}
\end{theorem}
The symbol $\lrcorner$ denotes the obvious contraction map between $S$
and $D$. We refer to \cite{AG:McExt} for a more detailed explanation.

One can then show the following:
\begin{theorem}
The map $d_{CD}: C\otimes_B D \to C\otimes_B D$ defined above
satisfies
\begin{equation}
  d^2_{CD} = -f.\id_{C\otimes D},  \label{eq:d2CD}
\end{equation}
where
\begin{equation}
  f = \sum_{i=1}^c f_i X_i.
\end{equation}
\end{theorem}
This is also proven in \cite{AG:McExt}.

Most importantly we have the following, again copied from
\cite{AG:McExt}:
\begin{theorem}
  Set $D'=D\otimes_B A$ and $y'=y\otimes 1$ for $y\in D$. The map
\begin{equation}
  \partial:C\otimes_B D'\to C\otimes_B D',
\end{equation}
given by
\begin{equation}
  \partial(x\otimes y') = \sum_{\gamma\in \NN^c} d_\gamma(x)\otimes 
     (X^\gamma\lrcorner\,y)'
\end{equation}
is an $A$-linear map of homological degree $-1$. This, together with the map
\begin{equation}
  q':C\otimes_B D'\to M,
\end{equation}
given by
\begin{equation}
q'(x\otimes y') = \begin{cases} y.r(x)&\text if \deg(y')=0\\
   0&otherwise\end{cases}
\end{equation}
is a resolution of $M$ by free $A$-modules.
\end{theorem}
Note that the map $r$ refers to the one in (\ref{eq:Cres}) and
$\deg(y')$ refers to the homological degree.

This yields a very practical construction of the free resolution of
$M$ that we will need. In particular this algorithm can be efficiently
implemented in Macaulay 2 as described in \cite{AG:McExt}.

Note that such a free resolution is always
infinitely long since $D$ has unbounded homological degree. However,
$\partial$ is represented by a finite-dimensional matrix with
entries in the ring $S$.

Given such a resolution of $M$, it is easy to compute the groups
$\Ext_A(M,N)$ for any $A$-module $N$. Note that the action of $S$ on
the free resolution descends to $\Ext_A(M,N)$ and thus $\Ext_A(M,N)$
is an $S$-module. Again $\bigoplus_k\Ext_A^k(M,N)$ can typically be
infinite-dimensional but it can always be presented as a finitely-generated
$S$-module. The variables $X_i$ yield maps
\begin{equation}
  X_i:\Ext_A^j(M,N)\to \Ext_A^{j+2}(M,N).
\end{equation}
As such, the homological degree we have defined is {\em negative\/} to
what one might normally use in the derived category. This is not
surprising since we graded (\ref{eq:Cres}) according to homology
rather than cohomology.

Unfortunately it is not these Ext groups that compute the Hilbert
spaces of open strings.  Having said that, the construction is
extremely relevant, as we now describe.

\subsection{A Quotient Category} \label{s:quo}

For the time being, let us restrict to the case of a hypersurface,
$c=1$. It is tempting to note the similarity between (\ref{eq:mfact2})
and (\ref{eq:d2CD}). The only difference is the appearance of the
extra factor $-X_1$ in (\ref{eq:d2CD}). We will now show that we may
convert the construction of section \ref{s:resn} to that required in
section \ref{ss:Dbrane} simply by setting $X_1$ equal to $-1$.

Let us also restrict to the case where $A$ is singly-graded and $f_1=W$
has degree $d$. The homological and ring grading of $X_1$ is then
equal to $(-2,-d)$. The only way we could set $X_1$ equal to $-1$ is if
we identify the homological and ring grading accordingly. Indeed we
would require $M[2]=M(d)$ exactly like in (\ref{eq:2=d}). This is
further evidence that we are on the right track.

In order to prove our assertion we need to introduce various
categories and consider functors between them.

Let $\Ch(\text{gr-}A)$ be the category whose objects are bounded chain
complexes of graded $A$-modules and whose morphisms are chain maps of
degree 0. Recall that GrPair($W$) is the category of matrix
factorizations (without modding out by homotopy) defined in section
\ref{ss:Dbrane}. We attempt to define a functor from $\Ch(\text{gr-}A)$ to
GrPair($W$) as follows:
\begin{itemize}
\item Any object $M^\bullet$ in $\Ch(\text{gr-}A)$ can be represented
by a bounded complex of $B$-modules annihilated by $W$. 
\item This complex
may then be represented by a finite free resolution, i.e., a finite
complex $C^\bullet$ of free $B$-modules. In order words, we have a
chain map $C^\bullet\to M^\bullet$ which is a quasi-isomorphism.
\item Now take this chain complex $C^\bullet$ and form the direct sum of its
elements to form $C$ as in (\ref{eq:C}) as well as the differential
$d_C:C\to C$.
\item Then apply the construction of theorem \ref{th:mf} to form a map
  $p:C\to C$ where
\begin{equation}
  p(x) = \sum_{\gamma\in\NN} (-1)^\gamma d_\gamma(x).
\end{equation}
This is essentially the same as (\ref{eq:dCD}) where $D$ has been
collapsed to $B$ (with an alternating sign) and thus we have
effectively set $X_1=-1$. The relation (\ref{eq:d2CD}) becomes
$p^2=W$.
\item Decomposing $C$ by homological degree we put
  $P_0=C^{\mathrm{even}}$ and $P_1=C^{\mathrm{odd}}$, and $p$
  decomposes into two maps $p_0$ and $p_1$ between these modules. Thus
  we obtain an object in GrPair($W$).
\end{itemize}

The only problem with the above construction is that we required a
choice of free resolution $C^\bullet$. Naturally the solution to this
problem is to go to the homotopy category. It is easy to show that a
chain homotopy in $\Ch(\text{gr-}A)$ will be mapped to a homotopy in
GrPair($W$) by the above construction. 
Let $K(\text{gr-}A)$ denote the category whose objects are
complexes of graded free $A$-modules and whose morphisms are chain
maps modulo chain homotopies. We also impose the following finiteness
condition of $K(\text{gr-}A)$. The complexes are bounded to the right
but may be infinite to the left. However they can be finitely
represented by $S$-modules as in section \ref{s:resn}.

This yields a well-defined functor
\begin{equation}
  K(\text{gr-}A) \to \mathrm{DGr}B(W).
\end{equation}

We now claim that the category $K(\text{gr-}A)$ is equivalent to
$\DC(\text{gr-}A)$, the bounded derived category of $A$-modules.  This
may be proven by using, for example, the result of section 3.10 of
\cite{GM:Hom} combined with the fact that any bounded complex of
$A$-modules has a free resolution satisfying the above finiteness
condition on $K(\text{gr-}A)$.

This yields a functor
\begin{equation}
  G':\DC(\text{gr-}A) \to \mathrm{DGr}B(W). \label{eq:F}
\end{equation}
One may check that this is an exact functor between triangulated
categories. 

For the next step we need to recall the definition of a quotient
triangulated category (see \cite{Orlov:LG}, for example). 
Given a triangulated category
$\mathsf{D}$ and full triangulated subcategory $\mathsf{N}$, we
define the quotient $\mathsf{D}/\mathsf{N}$ as follows. The objects in 
$\mathsf{D}/\mathsf{N}$ are the same as the objects in $\mathsf{D}$.
Consider the set of morphisms $\Sigma$ in $\mathsf{D}$ whose mapping
cones lie in $\mathsf{N}$. In other words $f:\mathsf{a}\to\mathsf{b}$
lies in $\Sigma$ if and only if we have a distinguished triangle
\begin{equation}
\xymatrix{
\mathsf{a}\ar[rr]^f&&\mathsf{b}\ar[dl]\\
&\mathsf{n}\ar[ul]|{[1]}
}  \label{eq:quodef}
\end{equation}
where $\mathsf{n}$ is an object in $\mathsf{N}$. The octahedral axiom
can be used to show that $\Sigma$ is multiplicatively closed.  The
morphisms in $\mathsf{D}/\mathsf{N}$ are then defined by
``localizing'' on the set $\Sigma$. That is, we invert the elements of
$\Sigma$ in the same way that quasi-isomorphisms are inverted in
defining the derived category.

Note, in particular, that the zero map $0\to\mathsf{n}$ is in $\Sigma$
so that any element of $\mathsf{N}$ is isomorphic to zero in
$\mathsf{D}/\mathsf{N}$. 

Let $\mf{Perf}(A)$ be the full subcategory of $\DC(\text{gr-}A)$
consisting of objects which may be represented by finite-length
complexes of free $A$-modules of finite rank.  The category
$\DSing(A)$ is then defined as the quotient (in the above sense)
\begin{equation}
   \DSing(A) = \frac{\DC(\text{gr-}A)}{\mf{Perf}(A)}.
\end{equation}

$\DSing(A)$ has the following universal property. If
$U:\DC(\text{gr-}A) \to \mathcal{C}$ is an exact functor, for some
triangulated category $\mathcal{C}$, such that $U(M)\cong0$ for any
object $M$ in $\mf{Perf}(A)$, then $U$ factors through $\DSing(A)$.
  
The functor $G'$ in (\ref{eq:F}) indeed has the above property. Let us
consider applying $G'$ to the $A$-module $A$ itself. First we represent
$A$ by a free resolution of $B$-modules. This is the complex
\begin{equation}
\xymatrix@1{
  B(-d)\ar[r]^-W&B.
}\end{equation}
Thus $C=B(-d)\oplus B$. It is then easy to show that $d_1$ maps $B$ to
$B(-d)$ as a multiplication by $-1$ and $d_j=0$ for $j>1$. Thus
\begin{equation}
  p = d_0-d_1 = \left(\begin{matrix}0&1\\W&0\end{matrix}\right).
\end{equation}
That is, $A$ is mapped to the trivial matrix factorization
$W=W.1$. The identity map of this matrix factorization to itself is a
null-homotopy. Thus, this object is isomorphic to zero in
$\mathrm{DGr}B(W)$. That is, $G'(A)\cong0$. Since $G'$ is an exact
functor, it follows immediately that $G'$ applied to any finite complex
of finitely-generated free $A$-modules yields zero. We have therefore defined
an induced functor
\begin{equation}
  G:\DSing(A)\to \mathrm{DGr}B(W).
\end{equation}

This functor is actually an equivalence of categories. To see this one
can construct an inverse functor as follows. Consider the functor from
GrPair($W$) to the category of graded $A$-modules given by the map
$\operatorname{Coker}(p_1)$. As shown in \cite{Orlov:mfc} this induces
a functor
\begin{equation}
  F:\mathrm{DGr}B(W)\to \DSing(A).
\end{equation}
Now consider an object $\bar P$ in $\mathrm{DGr}B$. Applying $GF$ to
$\bar P$ will produce another matrix factorization which is typically
larger than $\bar P$. However, both $\bar P$ and $GF(\bar P)$
represent free resolutions of the same module and so are equivalent in
the homotopy category $\mathrm{DGr}B(W)$. Thus $GF\cong\id$. Similarly
applying $FG$ to a complex of modules yields a complex of modules that
differs from the original by a finite complex of free modules and
hence $FG\cong\id$. Essentially, by Eisenbud's construction
\cite{Eis:mf} the original complex has a free resolution which will
{\em eventually\/} become 2-periodic to the left, whereas $FG$ applied to this
complex has a free resolution that is {\em immediately\/} the same 2-periodic
resolution. 

We have therefore proven
\begin{theorem} \label{th:Geq}
  The functor 
\begin{equation}
  G:\DSing(A)\cong \mathrm{DGr}B(W)
\end{equation}
is an equivalence of triangulated categories.
\end{theorem}

That these two categories are equivalent is not a new result --- it
was proven in \cite{Orlov:mfc}. What is useful to us is that the
equivalence is induced by the explicit functor $G$.

What we have shown is that the quotient $\DC(\text{gr-}A)
\to \DSing(A)$ is essentially performed by setting $X_1=-1$. To be more
precise, let $M^\bullet$ and $N^\bullet$ be two objects in
$\DC(\text{gr-}A)$. As shown in \cite{AG:McExt} and section
\ref{s:resn}, the group 
$\bigoplus_{j,n}\Hom_{\DC(\text{gr-}A)}(M^\bullet, N^\bullet[j](n))$ is a
finitely-generated $S$-module. Let $B$ be regarded as an $S$-module
where the action of $X_1$ is multiplication by $-1$. Theorem \ref{th:Geq} shows
that
\begin{equation}
  \bigoplus_{j,n}\Hom_{\DSing(A)}(M^\bullet, N^\bullet[j](n)) \cong
  \bigoplus_{j,n}\Hom_{\DC(\text{gr-}A)}(M^\bullet, N^\bullet[j](n))
\otimes_S B.  \label{eq:Homq}
\end{equation}

\subsection{A 2-Brane on a twisted cubic} \label{ss:cubic}

To clarify the construction of the previous section and to show how to
practically implement the procedure we give an example using Macaulay
2. 

We will give an example that is fairly simple so that we do not need
to do any extra programming. Having said that, it is sufficiently
complicated that computations ``by hand'' would be fairly awkward.

Let $X$ be the quintic 3-fold defined by the equation
\begin{equation}
W = {{x}}_{0}^{3} {{x}}_{1} {{x}}_{{2}}+{{x}}_{1}^{3} {{x}}_{{2}}^{2}+
{{x}}_{1}^{2} {{x}}_{{2}}^{3}-{{x}}_{0} {{x}}_{{2}}^{4}-{{x}}_{0}^{4} 
{{x}}_{{3}}-{{x}}_{1}^{4} {{x}}_{{3}}+{{x}}_{{3}}^{4} {{x}}_{{4}}+{{x}}_{{4}}^{5}.
\end{equation}
We would like to consider D-Branes wrapping the ``twisted'' cubic
rational curve defined by the ideal
\begin{equation}
  I = (x_1^2-x_0x_2,x_2^2-x_1x_3,x_1x_2-x_0x_3,x_4).
\end{equation}
We will compute the dimensions of the Hilbert spaces of open strings
beginning and ending on this D-brane, together with certain ``twists''
of the D-brane.

Let $B=k[x_0,\ldots,x_4]$ and let $A=B/(W)$ as above. Then $X=\Proj
A$. The sheaf supported on the cubic curve is then associated to the
$A$-module $M=A/I$.

Usually, given a sheaf (or complex of sheaves) in the geometric
picture of D-branes, one needs to go through a nontrivial process to
obtain an $A$-module in $\DSing(A)$ that represents the same D-brane
in the \LG\ picture. The precise algorithm was given in
\cite{me:lgdict}. Having said that, it was shown in \cite{me:lgdict}
that this process is trivial for projectively normal rational curves.
Thus the $A$-module $M$ correctly represents our 2-brane in
$\DSing(A)$.

Let us proceed with our example in Macaulay 2. For the first few lines
of input we will suppress printing the output. What follows is the
basic setup defining the rings and modules above. Note that, as usual
in Macaulay, we let $k$ be a finite field rather than the complex
numbers to improve the efficiency of the computation. As long as there
are no unfortunate coincidences between the coefficients in the
polynomials that Macaulay manipulates and the characteristic of $k$,
this should not affect the results.

{\small
\begin{verbatim}
i1 : kk = ZZ/31469
i2 : B = kk[x_0..x_4]
i3 : W = x_0^3*x_1*x_2+x_1^3*x_2^2+x_1^2*x_2^3-x_0*x_2^4-x_0^4*x_3-
         x_1^4*x_3+x_3^4*x_4+x_4^5
i4 : A = B/(W)
i5 : M = coker matrix{{x_1^2-x_0*x_2, x_2^2-x_1*x_3, x_1*x_2-x_0*x_3, x_4}}
\end{verbatim}}

Now we use the internal Macaulay routine described in \cite{AG:McExt}
to compute the $S$-module $\Ext_A^*(M,M)$:

{\small
\begin{verbatim}
i6 : ext = Ext(M,M)

o6 = cokernel {-2, 1} | X_1 0   0      0      0              
              {-2, 1} | 0   X_1 0      0      0              
              {-1, 1} | 0   0   0      0      0              
              {-1, 1} | 0   0   0      0      0              
              {-1, 1} | 0   0   0      0      -X_1x_3^2      
              {-1, 1} | 0   0   0      0      -X_1x_3^2      ...
              {-1, 1} | 0   0   X_1x_1 X_1x_0 -15734X_1x_3^2 
              {-1, 1} | 0   0   0      0      X_1x_3^2       
              {0, 0}  | 0   0   0      0      0              

                                                                        9
o6 : kk [X ,x ,x ,x ,x ,x ]-module, quotient of (kk [X, x ,x ,x ,x ,x ])
          1  0  1  2  3  4                            1  0  1  2  3  4

\end{verbatim}}
The output above is quite large and we have suppressed most of it. The
first column of the output represents the bi-degrees of the generators
of this module. The first degree is the homological degree discussed
in the previous section and the second degree is the original degree
associated to our graded ring $B$.
For example, the generator associated to the first row
is an element of $\Ext^2_A(M,M(1))$. The final row represents the
identity map. The 44 columns of the matrix in the rest of the output
represent relations between these generators. 

Next we need to pass to the quotient category $\DSing(A)$ by setting
$X_1=-1$. When we do this, we will collapse the bigrading to a single
grading satisfying $M[2]=M(5)$. To keep track of this single grading
we map an old grade of $(a,b)$ to the single grade $-5a+2b$. Thus the
degrees of $x_0,\ldots,x_4$ in our new ring will be 2. We denote this
new ring {\tt B2}.

The following code sets {\tt pr} equal to the map whose cokernel
defines $\Ext^*_A(M,M)$ above and we define our rings $S$ and {\tt B2}.
{\small
\begin{verbatim}
i7 : pr = presentation ext
i8 : S = ring target pr
i9 : B2 = kk[x_0..x_4,Degrees=>{5:2}]
i10: toB = map(B2, S, {-1,x_0,x_1,x_2,x_3,x_4}, 
                      DegreeMap => ( i -> {-5*i#0+2*i#1}))
\end{verbatim}}
The last line above is the heart of our algorithm. It defines a ring map
which sets $X_1=-1$ and defines how we map degrees. It is now simple to
compute the Ext's in the quotient category $\DSing(A)$ by constructing
the tensor product as in (\ref{eq:Homq}):
{\small
\begin{verbatim}
i11 : extq = prune coker(toB ** pr)

o11 = cokernel {7} | 0   0            0             0            0     
               {7} | 0   0            0             0            0     
               {7} | 0   0            0             0            0     
               {7} | 0   0            0             0            0     ...
               {7} | 0   0            0             0            0     
               {7} | 0   0            0             0            0     
               {0} | x_4 x_2^2-x_1x_3 x_1x_2-x_0x_3 x_1^2-x_0x_2 x_3^4 

                               7
o11 : B2-module, quotient of B2

\end{verbatim}}
Finally we may compute the dimensions of Hilbert spaces of open
strings by computing the dimensions of the above module at specific
degrees. This, of course, is the Hilbert function.
{\small
\begin{verbatim}
i12 : apply(20, i -> hilbertFunction(i, extq))

o12 = {1, 0, 4, 0, 7, 0, 10, 6, 6, 10, 0, 7, 0, 4, 0, 1, 0, 0, 0, 0}

o12 : List
\end{verbatim}}
The above list represents the dimensions of
$\Hom_{\DSing(A)}(M,M\langle i\rangle))$ for $i=0\ldots19$, where we
use angle brackets to represent twisting with respect to the single
grading we have defined. That is,
\begin{equation}
  \Ext^k(M,M(r)) = \Hom_{\DSing(A)}(M,M\langle 5k+2r\rangle).
\end{equation}
Note that Serre duality implies $\Hom_{\DSing(A)}(M,M\langle
i\rangle))\cong\Hom_{\DSing(A)}(M,M\langle 15-i\rangle))$ which is
consistent with the above output. For open strings beginning and
ending on the same untwisted 2-brane $M$ we immediately see
\begin{equation}
\begin{split}
  \Ext^0(M,M)=\C,\qquad&\Ext^1(M,M)=0\\
  \Ext^2(M,M)=0,\qquad&\Ext^3(M,M)=\C.
\end{split}
\end{equation}
This shows that our twisted cubic curve has normal bundle
$\O(-1)\oplus\O(-1)$.

We could also compute open string Hilbert spaces between $M$ and
twists of $M$. (Physically this twisting corresponds to monodromy
around the \LG\ limit as discussed in \cite{me:lgdict}.) For example,
$\Ext^1(M,M(1))=\C^6$.


\section{Superpotentials}  \label{s:superP}

The derived category associated to topological D-branes is endowed
with an $A_\infty$-structure as discussed in
\cite{KS:mir,Laz:super,Toms:Ainf,DGJT:D,HLW:Ainf,AK:ainf}.  This
$A_\infty$-structure also manifests itself as an effective
superpotential in the D-Brane world-volume theory. Viewing D-branes
in the geometric phase as complexes of coherent sheaves, this
$A_\infty$-structure can be quite tricky to compute. It is associated
with obstructions to deformations of the associated geometric objects.

Now that we have a much more computationally favourable setting in the
\LG\ phase, we should have an easier time discovering the
$A_\infty$-structure and thus computing superpotentials. Let us review
how we perform this computation.

Let $P$ denote $P_0\oplus P_1$. Our D-brane is a map $p:P\to P$
satisfying $p^2=W$. Now consider another map $a:P\to P$ which has a
definite parity, i.e., the map acts within $P_0$ and $P_1$ or it
exchanges them. (The literature such as \cite{KL:TLG1} often calls
even parity maps ``bosonic'' and odd parity maps ``fermionic''.) We
denote the parity $(-1)^a$. Now define a differential
\begin{equation}
  da = ap - (-1)^apa.  \label{eq:d}
\end{equation}
It follows that $d^2=0$ and the cohomology of this operator is exactly
the open string Hilbert spaces we discussed in section
\ref{ss:Dbrane}.

We also have a simple composition of maps. That is, if $a$ and $b$ are
both maps $P\to P$ then so is $ab$. This composition satisfies the
Leibniz rule with respect to the differential. Thus we have the
structure of a $\Z_2$-graded differential graded algebra. For the \LG\
orbifold we consider graded modules as in section \ref{ss:Dbrane}. As
before, this can be used to extend the $\Z_2$-grading to a
$\Z$-grading. (In the language of homological algebra we are defining
the ``Yoneda product'' between the corresponding $\Ext$ groups.)

As explained in \cite{Kel:Ainf,AK:ainf}, for example, a differential
graded algebra gives rise to an $A_\infty$-structure on the cohomology
of this algebra which is unique up to an $A_\infty$-isomorphism thanks
to Kadeishvili's theorem \cite{Kad:Ainf} as we now review.

An $A_\infty$-algebra $\cA$ comes equipped with a set of products
\begin{equation}
  m_k: \cA^{\otimes k}\to\cA,
\end{equation}
satisfying particular relations (see \cite{Kel:Ainf}, for example). An
$A_\infty$-morphism from $\cA$ to $\cB$ is a collection of maps 
$f_k: \cA^{\otimes k} \to\cB$ satisfying
\begin{equation}
  \sum_{r+s+t=n}(-1)^{r+st} f_u(\id^{\otimes r}\otimes m_s\otimes \id^{\otimes
  t}) =
  \sum_{\genfrac{}{}{0pt}{}{1\leq r\leq n}{i_1+\ldots+i_r=n}}
     \!\!(-1)^q m_r(f_{i_1}\otimes
  f_{i_2}\otimes\cdots\otimes f_{i_r}), \label{eq:Amorph}
\end{equation}
for any $n>0$ and $u=n+1-s$. The sign on the right is given by
\begin{equation}
  q = (r-1)(i_1-1) + (r-2)(i_2-1) + \ldots + (i_{r-1}-1).
\end{equation}
A differential graded algebra $\cB$ is trivially an $A_\infty$-algebra
where $m_1$ is the differential, $m_2$ is the product and all higher
$m_k$'s vanish. Let $H(\cB)$ denote the cohomology of $\cB$ and choose
an embedding $i:H(\cB)\to\cB$ which we set equal to $f_1$. One may
then iteratively apply (\ref{eq:Amorph}) for increasing values of $n$
to define higher $f_k$'s and $m_k$'s on $H(\cB)$ to define an
$A_\infty$-structure on $H(\cB)$. As stated above, this can be shown
to be unique up to an $A_\infty$-isomorphism.

Given a topological field theory of open strings we may use the BRST
operator $Q$ as the differential and the joining of open strings as
the product. The resulting $A_\infty$-structure on the Hilbert spaces
of open strings then yields the desired one associated to the
superpotential, at least up to an $A_\infty$-isomorphism. This
ambiguity seems to be unavoidable without using information beyond the
topological field theory.

Anyway, we are clearly in a position to compute the
$A_\infty$-structure. We simply need a representative map $P\to P$ for
each basis element of $\Ext^k(P,P)$. Unfortuantely, a
na\"\i ve approach to the problem can easily come up against
computational limits.

Suppose we consider a D-brane such that $C$, as constructed in section
\ref{s:resn} is of rank $N$. Thus $p$ is an $N\times N$ matrix. Now
consider the space of all homomorphisms $C\to C$. Such homomorphisms
are also $N\times N$ matrices and thus we have an
$N^2$-dimensional space of Hom's. It follows that the differential $d$
in (\ref{eq:d}) is represented by an $N^2\times N^2$ matrix. Since $N$
can easily be of the order of 100 for relatively simple D-branes, we
see that this direct method can be very unwieldy.

This is, of course, not how Macaulay computed the Ext groups in the
last example. In order to compute $\Ext_A(M,M)$, we need to find a
resolution $C$ of $M$ and then compute the cohomology of $\Hom(C,M)$
which is just an $N\times N$ matrix. Thus, Macaulay would represent an
element of $\Ext_A(M,M)$ by an element of $\Hom(C,M)$. Because of the
properties of projective modules, we may lift\footnote{There is a
  subtlety here. In order to lift this map systematically we require
  $d_C$ to have a strictly negative homological degree. This is true
  in the category $\DC(\textrm{gr-}A)$ but is no longer true when we
  set $X_1=-1$. Thus we find representatives of our open strings as
  matrices {\em before\/} we set $X_1=-1$.} to the desired map from
$C$ to $C$:
\begin{equation}
\xymatrix{C\ar[r]\ar[rd]^-\alpha\ar@{-->}[d]^-{\alpha'}&M\\
  C\ar[r]&M.}
\end{equation}
If we are considering an object in the derived category that cannot be
represented by a single $A$-module $M$, then we may apply the mapping
cone construction to build the necessary map.

Let us illustrate the procedure for a very simple example. Actually
this example has already been analyzed in \cite{ADED:obs} where the
computational complexity was reduced by using tensor product
structures rather than the approach we use here. Our method can be
applied to any D-brane but we will stick to this easy case for the
reader's benefit. Here we need to do a little more Macaulay programing as
we need to get ``inside'' the code for Ext listed in \cite{AG:McExt}
to find the map $p$. We spare the reader the details.

Let $B=k[x_0,\ldots,x_4]$, $A=B/(W)$ and $X=\Proj A$ be the Fermat quintic
given by
\begin{equation}
  W = x_0^5+x_1^5+x_2^5+x_3^5+x_4^5.
\end{equation}
Again we will consider a rational curve to avoid the complexities of
mapping between $\DC(X)$ and $\DSing(A)$. The one we choose is given
by $x_0+x_1=x_2+x_3=x_4=0$. That is, our 2-brane is given by the $A$-module
\begin{equation}
  M = \frac A{(x_0+x_1,x_2+x_3,x_4)}.
\end{equation}

The map
$p$ is given by the following matrix:
\begin{equation}
  p = \left(\begin{matrix}0&a_0&a_1&a_2&0&0&0&0\\
b_0&0&0&0&-a_1&-a_2&0&0\\
b_1&0&0&0&a_0&0&-a_2&0\\
b_2&0&0&0&0&a_0&a_1&0\\
0&-b_1&b_0&0&0&0&0&a_2\\
0&-b_2&0&b_0&0&0&0&-a_1\\
0&0&-b_2&b_1&0&0&0&a_0\\
0&0&0&0&b_2&-b_1&b_0&0
\end{matrix}\right),
\end{equation}
where $a_0=x_0+x_1$, $a_1=x_2+x_3$, $a_2=x_4$,
$b_0=x_0^4-x_0^3x_1+x_0^2x_1^2-x_0x_1^3+x_1^4$,
$b_1=x_2^4-x_2^3x_3+x_2^2x_3^2-x_2x_3^3+x_3^4$ and $b_2=x_4^4$.

Copying the procedure in section \ref{ss:cubic} we obtain a
particularly easy answer for $\Ext(M,M)$. There are only two
generators. One is given by the identity map $C\to C$ so we will call
it $\id$. The other generator has grade 3 and we call it $g$. $g$ is
easily determined to be
\begin{equation}
  g = \left(\begin{matrix}0&0&0&1&0&0&0&0\\
0&0&0&0&0&-1&0&0\\
0&0&0&0&0&0&-1&0\\
-x_4^3&0&0&0&0&0&0&0\\
0&0&0&0&0&0&0&1\\
0&x_4^3&0&0&0&0&0&0\\
0&0&x_4^3&0&0&0&0&0\\
0&0&0&0&-x_4^3&0&0&0\end{matrix}\right).
\end{equation}

Macaulay tells us that, for the $B$-module $\Ext(M,M)$ both
generators $\id$ and $g$ are independently annihilated by the ideal
\begin{equation}
  I = (x_0+x_1,x_2+x_3,x_4,x_0^4,x_2^4).  \label{eq:I1}
\end{equation}
This immediately gives us the following:
\begin{equation}
\begin{alignedat}{2}
  \Ext^0(M,M)&=\C &\quad&\textrm{generated by $\id$}\\
  \Ext^1(M,M)&=\C^2& &\textrm{generated by $x_0g$ and $x_2g$}\\
  \Ext^2(M,M)&=\C^2& &\textrm{generated by $x_0^2x_2^3\id$ and 
              $x_0^3x_2^2\id$}\\
  \Ext^3(M,M)&=\C& &\textrm{generated by $x_0^3x_2^3g$.}
\end{alignedat}  \label{eq:31spec}
\end{equation}

We can now build the $A_\infty$-structure. It is easy to prove that
$\id$ acts as the identity for $m_2$ and that $m_k$ vanishes for $k>2$
if any of the entries are equal to $\id$. So the only
$A_\infty$-products we need to compute are $m_k(g,g,\ldots,g)$ for all
$k$.

Putting $n=2$ in (\ref{eq:Amorph}) gives
\begin{equation}
  im_2(g,g) = df_2(g,g) + g.g.
\end{equation}
But $g^2=-x_4^3\id$. Since $\id$ is killed by $x_4$ in $\Ext(M,M)$, we
see that $m_2(g,g)=0$. Saying that $\id$ is killed by $x_4$ means
that $x_4\id$ is a null-homotopic map. It is again straight-forward
using the same techniques we have already employed to compute this
null-homotopy. We find $x_4\id=qp+pq$, where
\begin{equation}
q = \left(\begin{matrix}0&0&0&0&0&0&0&0\\
0&0&0&0&0&0&0&0\\
0&0&0&0&0&0&0&0\\
1&0&0&0&0&0&0&0\\
0&0&0&0&0&0&0&0\\
0&-1&0&0&0&0&0&0\\
0&0&-1&0&0&0&0&0\\
0&0&0&0&1&0&0&0\end{matrix}\right).
\end{equation}
Thus $f_2(g,g)=x_4^2q$. At the next step $n=3$ yields\footnote{$f_2$
  is an odd degree operator so $(f_1\otimes f_2)(g\otimes
  g)=-f_1(g)\otimes f_2(g)$.}
\begin{equation}
  im_3(g,g,g) = -g\cdot f_2(g,g)
  -f_2(g,g)\cdot g + d f_3(g,g,g). \label{eq:m3}
\end{equation}
But $qg+gq=\id$, which implies $m_3(g,g,g)=0$ and
$f_3(g,g,g)=x_4q$. Similarly $m_4(g,g,g,g)=0$ and
$f_4(g,g,g,g)=-q$. For $n=5$ we obtain
\begin{equation}
  im_5(g,g,g,g,g) = -f_1(g).f_4(g,g,g,g) -
  f_4(g,g,g,g).f_1(g)+df_5(g,g,g,g,g).
\end{equation}
This is not a null-homotopic map and we obtain 
\begin{equation}
  m_5(g,g,g,g,g) = \id,  \label{eq:m5}
\end{equation}
with $f_5(g,g,g,g,g)=0$. 

Finally we consider the inductive computation of $m_k(g,g,\ldots)$ for $k>5$. 
It is easy to show that $f_k(g,g,\ldots,\id,\ldots,g,g)=0$. Therefore
the only contribution to the left side of (\ref{eq:Amorph}) is
$im_n$. All terms of the form $f_k.f_l$ will vanish for $k+l>5$ on the
right-hand side. It follows that $m_k(g,g,\ldots)=0$ for all $k>5$.

We have found a very simple result. The only nonvanishing product in
the $A_\infty$-algebra is (\ref{eq:m5}). We may now apply the methods
discussed in \cite{AK:ainf} to compute the superpotential. Because
(\ref{eq:m5}) is the only non-vanishing product we will have a
superpotential that is purely sextic.  Let us use X and Y to denote
the two chiral superfields associated with the two $\Ext^1$'s in
(\ref{eq:31spec}). One obtains the exact result
\begin{equation}
  W_{\mathrm{4d}} = \Tr\left(\sum_{\sigma\in\mathfrak{S}_6}
   X_{\sigma(1)}X_{\sigma(2)}X_{\sigma(3)}Y_{\sigma(4)}Y_{\sigma(5)}Y_{\sigma(6)}
  \right),
\end{equation}
where $\mathfrak{S}_6$ is the symmetric group on 6 elements.
This is consistent with the moduli space analyzed in \cite{AK:lines}
and the form conjectured\footnote{We seem to have had a little more
  luck than the authors of \cite{ADED:obs} even though our
  computations are very similar. They did not obtain an exact
  sextic and needed to conjecture an $A_\infty$-isomorphism that
  would restore it to this form.} in \cite{ADED:obs}.

We illustrated this method of computing the superpotential for a very
easy example that only required relatively small $8\times 8$
matrices. Macaulay can cope with much more complicated cases 
where other methods, such as those in \cite{ADED:obs}, would be
impractical.

\subsection{The closed to open string map} \label{s:clos}

As a byproduct of the above analysis we find a very practical way of
computing the closed to open string maps of
\cite{Hofman:2000ce,Laz:oc}.

It was shown in \cite{KL:TLG1} that, as far as correlation functions
are concerned, there is no difference between a closed string
associated to the monomial $m$ and an open string operator of the form
$m.\id$. This map $m\mapsto m.\id$ therefore yields the closed to open
string map. This idea, and its consequences for the superpotential has
also been discussed in \cite{Hori:2004ja}.

We can immediately read this map from the Macaulay
computations of $\Ext(M,M)$. So long as $M$ is not isomorphic to zero,
the identity map will be an element of $\Ext^0(M,M)$ and thus is one of
the generators given by Macaulay. The relations on this generator
tell us exactly which monomials map to zero and each other under the
closed to open string map.

For example, in the example above, $\id$ is annihilated by $I$ given
by (\ref{eq:I1}). So the closed strings given by monomials
$x_0^3x_2^2$ and $x_1^3x_2^2$ would both map to the same nontrivial
element of $\Ext^2(M,M)$ while $x_0x_1x_2x_3x_4$ maps to zero.


\section{The Linear $\sigma$-model Interpretation}  \label{s:lin}

\subsection{A conjecture} \label{ss:conj}

The quotient construction of section \ref{s:quo} has a very natural
interpretation in terms of the gauged linear $\sigma$-model of
\cite{W:phase}. This will lead us to the idea that different phases of
the linear $\sigma$-model are associated to different, but equivalent,
presentations of the category of topological B-type D-branes. The idea
of using the gauged linear $\sigma$-model to analyze the phase
structure of D-branes and the resulting categorical equivalences has
also been explored recently in \cite{HHP:lects}.

Let us consider a linear $\sigma$-model with a collection of chiral
fields $\phi_1,\phi_2,\phi_3,\ldots\phi_N$. Let the gauge group be
$\GU(1)^m$. We also have a global $\GU(1)$ R-symmetry under which the
superpotential, $\mathcal{W}$, has charge $-2$. Thus, each chiral field has a
collection of $m+1$ charges. Set
\begin{equation}
  S = k[\phi_1,\phi_2,\ldots,\phi_N],
\end{equation}
which will be a multi-graded algebra with $m+1$ gradings.

The equations of motion of the associated Lagrangian
are divided into equations associated to the D-term and equations
associated to the F-term. The D-terms impose
a set of conditions that certain combinations of fields are not
allowed to vanish. These conditions can be extracted from the
triangulation  of the fan associated to the phase of the linear
$\sigma$-model as discussed in \cite{AGM:II}. 

The subset of $\C^N=\Spec S$ which is disallowed by the D-term equations of
motion may be expressed in terms of an ideal $J\subset S$. This is the same
ideal (which was denoted $B$ by Cox) that appears in \cite{Cox:}.

The category $\DC(\textrm{gr-}S)$ is triangulated and, therefore, has a
shift functor $[\ldots]$. In addition, there are $m+1$ twist functors which
cause a shift in the multigrading. These are normally denoted by
parentheses $(\ldots)$.

The category $\mathrm{DGr}S(\mathcal{W})$ is the homotopy category of matrix
factorizations of $\mathcal{W}$ over the ring $S$. Following
\cite{Orlov:mfc}, one may show that
\begin{equation}
  \mathrm{DGr}S(\mathcal{W})\cong 
      \frac{\DC(\textrm{gr-}S')}{\mf{Perf}(S')},  \label{eq:q8}
\end{equation}
where $S'=S/(\mathcal{W})$. 

The category $\mathrm{DGr}S(\mathcal{W})$ has the homological degree
identified with one of the gradings. This latter grading is the one we
identified with the R-charge. So $\mathrm{DGr}S(\mathcal{W})$ has one
shift functor $[\ldots]$ and $m$ twist functors.

We introduce the following notation. Let $T$ be an object in
$\DC(\textrm{gr-}S)$ which is annihilated by $\mathcal{W}$. 
Then $T$ is an object in $\DC(\textrm{gr-}S')$ and thus, by
(\ref{eq:q8}), an object in $\mathrm{DGr}S(\mathcal{W})$.
Consider the collection of all shifted and twisted objects
$T[\ldots](\ldots)$ and then find the minimal full triangulated
subcategory of $\mathrm{DGr}S(\mathcal{W})$ that contains this collection. We
denote the resulting subcategory $T^\triangle$. In other words,
$T^\triangle$ is formed by iteratively applying the mapping cone
construction to collections of $T$ and all of its translates and twists.

We now claim the following:
\begin{conjecture}
  In any given phase of the gauged linear $\sigma$-model, the category
  of topological B-type D-branes is given by the triangulated quotient
\begin{equation}
  \frac{\mathrm{DGr}S(\mathcal{W})}{T^\triangle},
\end{equation}
where $T=S/(J+(\mathcal{W}))$.
\label{cong:1}
\end{conjecture}

First we show that this conjecture is true for the quintic.
Here $S=[p,x_0,\ldots,x_4]$ and the superpotential may be chosen as
\begin{equation}
  \mathcal{W} = pW = p(x_0^5+x_1^5+x_2^5+x_3^5+x_4^5).
\end{equation}
We may choose the charges of our
superfields to be
\begin{equation}
\begin{array}{c|cccccc}
&p&x_0&x_1&x_2&x_3&x_4\\
\hline
R&-2&0&0&0&0&0\\
\GU(1)&-5&1&1&1&1&1
\end{array}
\end{equation}
This should look strikingly familiar to section \ref{s:resn}. This is
the bigrading structure given to the ring $S$ for the quintic where
$p$ plays the r\^ole of $X_1$.

As in earlier sections, we set $A=k[x_0,\ldots,x_4]/(W)$.  We now
claim that $\mathrm{DGr}S(\mathcal{W})$ is equivalent to the category
$\DC(\textrm{gr-}A)$. A functor $\DC(\textrm{gr-}A)\to
\mathrm{DGr}S(\mathcal{W})$ is essentially constructed in section
\ref{s:resn}. The inverse functor $\mathrm{DGr}S(\mathcal{W})\to
\DC(\textrm{gr-}A)$ can be constructed by taking a matrix
factorization and setting $p=0$ to obtain $d_o=d_C$ which yields a
free $B$-module resolution of an object in $\DC(\textrm{gr-}A)$.

The D-term equations depend on a parameter corresponding to the
complexified K\"ahler form.  If $X$ is ``small'', that is we are in
the \LG\ phase, then $p=0$ is excluded. That is, $J=J+(\mathcal{W})$
is the principal ideal $(p)$. The triangulated quotient is
particularly simple when $T$ is the quotient of $S$ by a principal
ideal as seen as follows. $S/(p)$ itself can be viewed as the cokernel
of a morphism which looks like multiplication by $p$. Similarly all
translates and shifts in grading of $S/(p)$ can be written as a
cokernel of multiplication by $p$. It follows that $T^\triangle$ in
this case consists of mapping cones of maps that involve
multiplication by $p$. These maps are localized simply by setting $p$
equal to a unit. 

Hence we perform the quotient in the conjecture simply by setting $p$
equal to some unit, say $-1$.  It is then clear that
$\mathrm{DGr}S(\mathcal{W})$ becomes $\mathrm{DGr}B(W)$ under this
quotient. So the conjecture is correct in the \LG\ phase.

In the other ``large radius limit'' phase, the D-terms demand that
$x_0,\ldots,x_4$ cannot simultaneously vanish, i.e.,
$J=(x_0,\ldots,x_4)$.  Now, forgetting about the $p$-action, $(S/J)$
is equal to $\textrm{tors-}A$, as discussed in \cite{me:lgdict}, for
example. Here $\textrm{tors-}A$ denotes ``torsion'' modules as defined
in \cite{Orlov:mfc}. It is then a result of Serre \cite{Serre:mp} that
the bounded derived category of coherent sheaves on $X$ is the
quotient $\DC(\textrm{gr-}A)/\textrm{tors-}A$ (see also
\cite{Orlov:mfc}). Thus the conjecture yields the desired result in
both phases for the quintic threefold.

This conjecture can also be shown to be true for some of the phases
associated to hypersurfaces in weighted projective spaces. This is
most easily demonstrated by an example. Consider the linear
$\sigma$-model with
\begin{equation}
  \mathcal{W}=p(x_0^4+x_1^4+x_2^4+t^4x_3^8+t^4x_4^8),
\end{equation}
and charges
\begin{equation}
\begin{array}{c|ccccccc}
&p&t&x_0&x_1&x_2&x_3&x_4\\
\hline
R&-2&0&0&0&0&0&0\\
Q_1&-8&0&2&2&2&1&1\\
Q_2&0&-2&0&0&0&1&1
\end{array}
\end{equation}

In the \LG\ phase $J=(p)(t)$ and the quotient in the conjecture sets
$p$ and $t$ to units. Setting $t$ equal to a constant removes the
$Q_2$-grading. Setting $p$ equal to $-1$ performs the quotient in
section \ref{s:quo}. Thus we end up with the category of matrix
factorizations of
\begin{equation}
  W = x_0^4+x_1^4+x_2^4+x_3^8+x_4^8,  \label{eq:2p}
\end{equation}
as expected.

In the ``orbifold'' phase $J=(t)(x_0,\ldots,x_4)$. We again set $t$
equal to a constant removing the $Q_2$-grading. This part of the quotient
leaves us with $\DC(\textrm{gr-}A)$ where $A=k[x_0,\ldots,x_4]/(W)$.
Then we divide out by the category of torsion modules. What
we obtain is the bounded derived category of coherent sheaves on the
hypersurface $W=0$ in the weighted projective stack
$\P^4_{\{2,2,2,1,1\}}$ as discussed in proposition 2.16 of
\cite{Orlov:mfc}. This is the expected result.

In the large radius phase the D-terms impose the constraints that
$(x_3,x_4)\neq(0,0)$ and $(x_0,x_1,x_2,t)\neq(0,0,0,0)$. The category
of modules we are left with is still doubly graded using both $Q_1$
and $Q_2$. This exactly reproduces Cox's description of the derived
category of coherent sheaves, as described in \cite{Cox:}, on the
hypersurface in the toric variety corresponding to the blow-up of the
above orbifold. Thus we again obtain the expected result.

The above three paragraphs easily generalize to the case of any
weighted projective space. The \LG, large radius, and (partially
resolved) orbifold phases are all in compliance with conjecture
\ref{cong:1}. 

What is new would be the so-called hybrid cases. In the case of
(\ref{eq:2p}) we have a hybrid $\P^1$ phase where the conditions
imposed are that $p=-1$ and that $(x_3,x_4)\neq(0,0)$. Imposing $p=-1$
will impose a quotient like in section \ref{s:quo} and therefore gives
us some kind of matrix factorization picture. Imposing
$(x_3,x_4)\neq(0,0)$ is similar to quotienting by torsion modules and
therefore has some of the character of coherent sheaves. Therefore, as
expected we have some kind of hybrid picture of geometry and \LG\
theories. We will not investigate this further here but it would be
interesting to work out further details.

\subsection{Complete Intersections}  \label{ss:CI}

As another application of conjecture \ref{cong:1} we may consider
cases of complete intersections in toric varieties which have a \LG\
phase. Consider, for example, the linear $\sigma$-model with a
superpotential
\begin{equation}
\mathcal{W} = p(x_0^3+x_1^3+x_2^3+x_3^3) +
q(x_0y_0^3+x_1y_1^3+x_2y_2^3),
\end{equation}
and charges
\begin{equation}
\begin{array}{c|ccccccccc}
&p&q&x_0&x_1&x_2&x_3&y_0&y_1&y_2\\
\hline
R&-2&-2&0&0&0&0&0&0&0\\
Q_1&-3&-1&1&1&1&1&0&0&0\\
Q_2&0&-3&0&0&0&0&1&1&1
\end{array}
\end{equation}
$S$ is the ring $k[p,q,x_0,x_1,x_2,x_3,y_0,y_1,y_2]$ with a tri-graded
structure given by these charges.

In the large radius phase we note that section \ref{s:resn} implies
that $\mathrm{DGr}S(\mathcal{W})$ is equivalent to
$\DC(\textrm{gr-}A)$ where
\begin{equation}
  A = \frac{k[x_0,x_1,x_2,x_3,y_0,y_1,y_2]}{(W_p,W_q)},
\end{equation}
and
\begin{equation}
\begin{split}
  W_p &= x_0^3+x_1^3+x_2^3+x_3^3\\
  W_q &= x_0y_0^3 + x_1y_1^3 + x_2y_2^3.
\end{split}
\end{equation}
The D-terms imply that $J$ is the ideal
$(x_0,x_1,x_2,x_3).(y_0,y_1,y_2)$. By conjecture \ref{cong:1} the
desired D-brane category is then
\begin{equation}
  \frac{\DC(\textrm{gr-}A)}{(A/J)^\triangle}.
\end{equation}
By Cox's construction this is
precisely the bounded derived category of coherent sheaves on $X$
where $X$ is Schimmrigk's manifold \cite{Schimmrigk:1987ke} defined by 
$W_p=W_q=0$ in $\P^3\times\P^2$.

In the \LG\ phase we set $p=q=-1$. This collapses the tri-graded
structure of $S$ to a single grading. Thus, if $M$ is an element of 
$\DC(\textrm{gr-}A)$ we now have the relation
\begin{equation}
  M[2] = M(3,0) = M(1,3).
\end{equation}
It is not immediately obvious how to generalize Orlov's constructions
\cite{Orlov:LG,Orlov:mfc} to the multiply-graded case and we will not
try to make any statements about $\DSing(A)$. Instead we note that we
can perform the necessary quotient by using the method of section
\ref{s:quo} and therefore we know how to compute the necessary Ext
groups by using Macaulay.\footnote{As of Macaulay 2 version 0.9.95 the
  necessary code for Ext is not written in a way that can handle
  multiple gradings. It is easy to rewrite the code in \cite{AG:McExt}
  such that it does. Future versions of Macaulay 2 are expected to be
  able to handle multiple gradings.}
It is also clear from (\ref{eq:d2CD}) that, by setting $p=q=-1$ (that
is, $X_1=X_2=-1$) we are forming matrix factorizations of $W_p+W_q$.
Thus, the category of D-branes in this \LG\ phase is again given by
matrix factorizations.

We should note that not any complete intersection can be analyzed
using matrix factorizations. This method is very much tied to the \LG\
picture and this phase may not exist in general. For example, the
intersection of two cubics in $\P^5$ has only a large radius \CY\
phase and a hybrid $\P^1$ phase. 


\section{Discussion}  \label{s:conc}

In this paper we have shown that the matrix factorization approach
to D-branes can be made quite practical for computations. Perhaps more
interestingly, we have also seen how the phase picture of the linear
$\sigma$-model naturally ties in with the derived category picture for
D-branes. There are a number of directions for further research that
present themselves.

In order to prove conjecture \ref{cong:1} we would need to carefully
analyze the boundary degrees of freedom in the linear $\sigma$-model
and show that one requires a matrix factorization of $\mathcal{W}$.
Then the conjecture should follow from imposing D-term conditions.
Note, in particular, that one does {\em not\/} need to prove the
appearance of the derived category directly by considering chain
complexes of boundary conditions.  There are analyses of linear
$\sigma$-models with boundaries in the literature at present, such as
\cite{Hori:lsm,lsm:multistep}, but these do not deal with matrix
factorizations. More interestingly, the recent work of
\cite{HHP:lects,HHP:linphase} does perform an analysis along these
lines and so promises a proof that we have a whole host of triangulated
categories that are equivalent to each other, generalizing the McKay
correspondence \cite{BKM:MisM} and Orlov's equivalence \cite{Orlov:mfc}.

Phases such as the hybrid models where one has a mixture of matrix
factorizations and coherent sheaves might well be worthy of study in
their own right. 

Another interesting idea associated to the different phases is that of
D-brane $\Pi$-stability. We know that $\Pi$-stability reduces to
$\mu$-stability (modulo some subtleties \cite{EL:mupi}) for the large
radius limit and that $\Pi$-stability reduces to $\theta$-stability
for orbifolds. $\Pi$-stability in the \LG\ phase is not as well
understood but some results were discussed in \cite{Wal:LGstab}. One
might expect some specific flavour of stability that can be associated
with each phase. This would be stated in terms of an abelian category
that perhaps would be the heart of a $t$-structure of the triangulated
category of D-branes in each case.

Clearly there are still many interesting properties of D-branes even
in the case of topological field theory that have yet to be fully
understood.


\section*{Acknowledgments}

I wish to thank S.~Katz, R.~Plesser and A.~Roy for useful discussions.
The author is supported by an NSF grant DMS--0606578.


\end{document}